\documentclass[lettersize,journal]{IEEEtran}
\usepackage{changes}
\usepackage[sort,numbers]{natbib}
\usepackage{amsmath,amsfonts}
\usepackage{algorithmic}
\usepackage{algorithm}
\usepackage{array}
\usepackage{caption}
\usepackage{amsthm, amssymb}

\usepackage{subcaption}
\usepackage{textcomp}
\usepackage{stfloats}
\usepackage{url}
\usepackage{verbatim}
\usepackage{graphicx}
\usepackage{color}
\usepackage{tikz}
\usepackage{booktabs}
\usepackage{makecell}
\def\BibTeX{{\rm B\kern-.05em{\sc i\kern-.025em b}\kern-.08em
    T\kern-.1667em\lower.7ex\hbox{E}\kern-.125emX}}
\usepackage{balance}
\begin{document}
\title{A Lightweight Authentication and Key Agreement Protocol Design for  FANET}
\author{\IEEEauthorblockN{Yao Wu$^{\ast}$, Ziye Jia$^{\ast}$, Qihui Wu$^{\ast}$,  and Yian Zhu$^{\ast}$  \\
 }\IEEEauthorblockA{$^{\ast}$The Key Laboratory of Dynamic Cognitive System of Electromagnetic Spectrum Space, Ministry of Industry and Information Technology, Nanjing University of Aeronautics and Astronautics, 211106, China. \\ E-mail: \{wu\_yao, jiaziye, wuqihui, zhuyian\}@nuaa.edu.cn }

\thanks{This work was supported  in part by National Natural Science Foundation of China under Grant 62231015, in part by National Natural Science Foundation of China under Grant 62301251, in part by the Aeronautical Science Foundation of China 2023Z071052007, in part by the Young Elite Scientists Sponsorship Program by CAST 2023QNRC001, and in part by the Postgraduate Research \& Practice Innovation Program of Jiangsu Province under Grant SJCX25\_0152. (\textit{Corresponding author: Ziye Jia}).}

 }


\maketitle
\thispagestyle{empty}
\begin{abstract}
The advancement of low-altitude intelligent networks enables unmanned aerial vehicle (UAV) interconnection via flying ad-hoc networks (FANETs), offering flexibility and decentralized coordination. However, resource constraints, dynamic topologies, and UAV operations in open environments present significant security and communication challenges. Existing multi-factor and public-key cryptography protocols are vulnerable due to their reliance on stored sensitive information, increasing the risk of exposure and compromise. This paper proposes a lightweight authentication and key agreement protocol for FANETs, integrating physical unclonable functions with dynamic credential management and lightweight cryptographic primitives. The protocol reduces computational and communication overhead while enhancing security. Security analysis confirms its resilience against various attacks, and comparative evaluations demonstrate its superiority in security, communication efficiency, and computational cost.

\end{abstract}


\section{Introduction}
\IEEEPARstart 
{U}{nmanned} aerial vehicles (UAVs)  are increasingly deployed in logistics, surveillance, and emergency response due to their operational efficiency and adaptability in challenging environments \cite{yang2023depth}. As UAV networks expand, the Internet of UAVs  has emerged to enable large-scale interconnection and coordination \cite{jia2025distributionally}, \cite{zhou2023aerospace}. However, the dynamic nature of UAV operations, where UAVs frequently join and leave the network, presents significant challenges for secure communication, network management, and reliable data transmission \cite{lu2024joint}. Specifically, the highly dynamic topology complicates network management, while the absence of fixed infrastructure limits the scalability and adaptability of conventional communication architectures.

Flying ad-hoc networks (FANETs) have emerged as a promising solution, enabling UAVs to autonomously form temporary networks without fixed infrastructure. This decentralized architecture enhances scalability and flexibility, making it well-suited for dynamic and resource-constrained environments. In FANETs, UAVs communicate directly to exchange position data, sensor readings, and mission parameters, supporting collaborative tasks. However, their open and decentralized nature introduces critical security risks \cite{11122503}. The absence of a central authority exposes FANETs to unauthorized access and data tampering \cite{pattaranantakul2023service}, while frequent topology changes complicate secure communication and authentication \cite{wu2024adaptive}. Moreover, UAVs operate in open and potentially hostile environments, making them vulnerable to physical capture attacks, which can lead to credential extraction and security breaches \cite{du2024tri}. These challenges necessitate an authentication and key agreement (AKA) protocol that is adaptive, lightweight, and resilient to physical capture attacks.

AKA protocols secure UAV communications by ensuring that only authorized entities participate in the network \cite{yaacoub2020security}. However, existing protocols often fail to address FANET-specific challenges. Many rely on computationally intensive cryptographic techniques, such as bilinear pairings and elliptic curve cryptography (ECC), which impose excessive computational and communication overhead, making them impractical for resource-constrained UAVs \cite{jia2024cooperative}. Additionally, multi-factor authentication methods often store sensitive credentials, increasing the risk of exposure in physical capture attacks. Moreover, existing protocols struggle with dynamic node management, failing to efficiently handle UAVs joining and leaving the network. These limitations underscore the need for a more efficient and secure AKA protocol tailored to the dynamic nature, resource constraints, and security threats of FANETs.

To address these issues, this paper proposes a lightweight AKA protocol leveraging physical unclonable functions (PUFs) for secure FANET communication. PUFs generate hardware-based device fingerprints without storing cryptographic secrets, mitigating physical capture risks in open environments \cite{liang2021mutual}. Unlike conventional PUF-based schemes that rely on pre-registered device relationships with centralized authorities, the proposed protocol integrates dynamic credential management with lightweight cryptographic primitives (e.g., hash functions and XOR operations). This design minimizes computational and communication overhead while adapting to FANET topology changes. By combining hardware-level security with efficient cryptographic techniques, the protocol achieves a balance among security, efficiency, and resource constraints. For clarity, the main contributions of this work are listed as follows.

\begin{itemize}
\item[$\bullet$]The proposed protocol ensures secure and efficient AKA in FANETs while addressing UAV resource constraints.

\item[$\bullet$]Security analysis demonstrates the resilience of the protocol against a wide range of known attacks.

\item[$\bullet$]Comparative analysis with existing protocols shows that the proposed protocol offers superior efficiency in security, functionality, communication, and computation overheads.

\end{itemize}

The remainder of the paper is organized as follows.  The system model is described in Section \ref{sec3}. Section \ref{sec4} introduces the proposed protocol. The security analysis of the proposed protocol is provided in Section \ref{sec5}. Performance comparisons are presented in Section \ref{sec6}. Finally, the conclusion is drawn in Section \ref{sec8}.

\section{System Model}
\label{sec3}

\begin{figure}[t]
\centering
\includegraphics[scale=0.4]{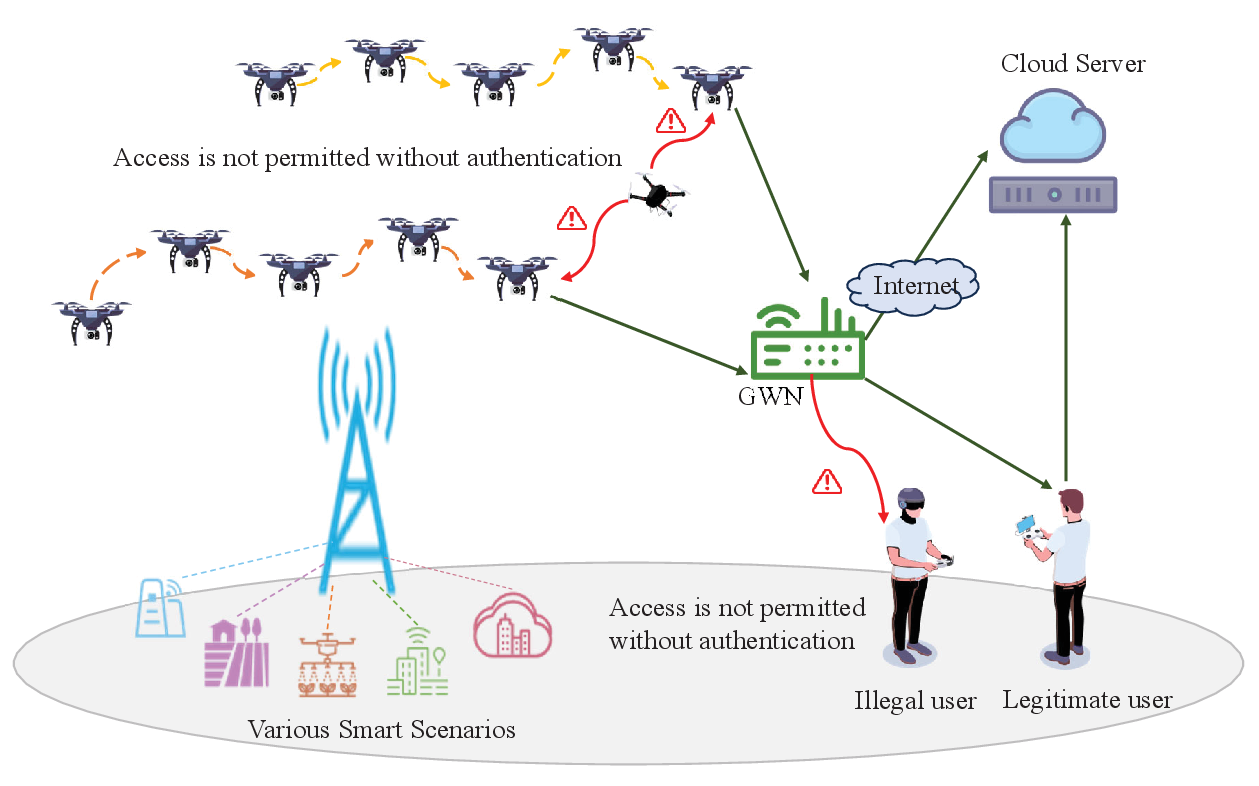}
\captionsetup{font={small},labelsep=period}
\caption{\label{system}\raggedright Identity authentication model for FANETs.} \vspace{-3mm}
\end{figure}

The FANET identity authentication model, illustrated in Fig. \ref{system}, consists of UAVs, base stations, a gateway node (GWN), users, and cloud servers. UAVs deployed in a target area form a FANET, enabling autonomous communication and collaboration for real-time environmental monitoring and task execution. UAVs collect and process mission-related data, transmitting it to the GWN via base stations. The GWN aggregates and forwards this data to the cloud server for analysis and storage. Users, with varying access privileges, require real-time data retrieval from UAVs. The FANET enforces role-based access control, ensuring operators can monitor mission progress while managers analyze data for decision-making. Authentication occurs as follows: a user submits login credentials to the GWN, which verifies the identity and forwards encrypted authentication data to the corresponding UAV. The UAV responds with key exchange information, establishing a session key for secure communications. Additionally, UAVs relay collected data to the GWN, which periodically uploads it to the cloud server. Pre-established encryption mechanisms protect data confidentiality and integrity. The cloud server, treated as a distinct identity node, undergoes a registration process similar to UAVs. Once authenticated, users gain access to stored data for post-mission analysis, supporting decision-making and operational optimization. To maintain network security, all UAVs and users undergo strict authentication, preventing unauthorized access and ensuring only legitimate entities interact with the system.

The proposed protocol is designed based on a comprehensive threat model that accounts for various security risks. Communication channels are assumed vulnerable to eavesdropping, message tampering, and deletion, as described by the Dolev-Yao model \cite{dolev1983security}. An adversary, denoted as $\mathcal{A}$, can intercept and analyze exchanged messages, potentially compromising sensitive information. Endpoint entities, including user devices and UAVs, may also be susceptible to compromise, allowing $\mathcal{A}$ to gain unauthorized access, manipulate data, or extract cryptographic keys and sensor readings. Furthermore, UAVs operating in hostile environments face physical security threats \cite{jia2025hierarchical}, such as capture or tampering, enabling $\mathcal{A}$ to exploit vulnerabilities in memory or firmware through power analysis or side-channel attacks \cite{messerges2002examining}. The GWN is assumed to be a trusted entity, resistant to compromise, as it plays a crucial role in maintaining the integrity and confidentiality of data exchanged within the network.

\begin{table}[t]
\centering
\captionsetup{justification=centering, labelformat=default,labelsep=newline,textfont=sc}
\caption{Key Notations}
\label{tab:symbols}
\begin{tabular}{@{}cl@{}}
\toprule
\textbf{Symbol} & \textbf{Description} \\ \midrule
$GWN$ & Gateway node (trusted authority)  \\ 
$ID_{G}$ & Identity of GWN \\
$U_i, UA_j$ & $i$th user and $j$th UAV, respectively \\
$ID_{i}, ID_{j}$ & Identity of $U_i$ and identity of $UA_j$, respectively \\
$PW_i$ & Password of $U_i$ \\
$SC_i$ & Smart card of $U_i$ \\
$BIO_i$ & Biometrics of $U_i$ \\
$s$ & Secret numbers of $GWN$ \\
$\sigma_i$ & Biometric secret key of $U_i$ \\ 
$\tau_i$ & Public reproduction threshold parameter of $U_i$ \\ 
$Gen(\cdot)$ & Fuzzy extractor generation procedure \\ 
$Rep(\cdot)$ & Fuzzy extractor reproduction procedure \\ 
$t$ & Error tolerance threshold used in fuzzy extractor \\
$n_i$, $n_j$, $n_k$ &  Random nonces of $U_i$, $GWN$ and $UA_j$, respectively \\
$TS_1,TS_2,TS_3$ & Current timestamps \\
$\Delta T$ & Maximum transmission delay associated with a message \\
$\|, \oplus$ & Concatenation and bitwise XOR, respectively \\
$h(\cdot)$ & Collision-resistant cryptographic one-way hash function \\
$\mathcal{A}$ & Adversary \\ \bottomrule
\end{tabular}
\end{table}

\section{Protocol Design}
\label{sec4}
The proposed protocol is a three-party authentication scheme based on PUF encryption, comprising six phases: the user registration, UAV registration, AKA, password and biometric update, smart card replacement, and dynamic UAV addition. This key notations are listed in Table \ref{tab:symbols}.

\subsection{User Registration}

\subsubsection{Step 1}
The user initiates registration by providing their unique identifier \( ID_i \), password \( PW_i \), and a randomly generated nonce \( n_i \). Using these, the user computes \( TID_i = h(ID_i \| n_i) \) and \( TPW_i = h(PW_i \| n_i) \), which are securely sent to the GWN.

\subsubsection{Step 2}
The GWN receives \( TID_i \) and \( TPW_i \), then combines them with its own identifier \( ID_G \) and a secret gateway parameter \( s \) to compute \( TC_{ID_i} = TID_i \oplus TPW_i \oplus h(ID_G \| s) \). This ensures the authentication process involves both the user’s credentials and the gateway’s identity. \( TC_{ID_i} \) is then sent back to the user securely.

\subsubsection{Step 3}
Upon receiving \( TC_{ID_i} \), the user uses their biometric data \( BIO_i \) to compute biometric parameters \( \sigma_i \) and \( \tau_i \) via \( Gen(BIO_i) \). The user then calculates \( A_i = n_i \oplus h(ID_i \| \sigma_i) \), \( B_i = h(ID_i \| TPW_i \| \sigma_i) \), and \( C_i = TC_{ID_i} \oplus TID_i \oplus TPW_i \). These values, along with other necessary information, are stored securely on the user’s smart card \( SC_i \).

\subsection{UAV Registration}

\subsubsection{Step 1}
The UAV selects a unique identifier \( ID_j \) and securely transmits it to the GWN to initiate the registration process.

\subsubsection{Step 2}
Upon receiving the UAV's identity, the GWN generates a random nonce \( n_j \) and computes the temporary identifier \( TID_j = h(ID_j \| n_j) \). It then generates the temporary certificate \( TC_{ID_j} = h(TID_j \| s) \) and produces a challenge \( C_j \), which are securely sent back to the UAV.

\subsubsection{Step 3}
The UAV computes a response \( R_j = PUF(C_j) \) using its PUF and sends it back to the GWN, while storing \( C_j \), \( ID_j \), and \( TC_{ID_j} \) for future authentication.

\subsubsection{Step 4}
The GWN securely stores the received data, including \( TC_{ID_j} \), nonce \( n_j \), and the challenge-response pair \( \langle C_j, R_j \rangle \), for future verification of the UAV's identity.

\begin{figure}[t]
\centering
\includegraphics[scale=0.483]{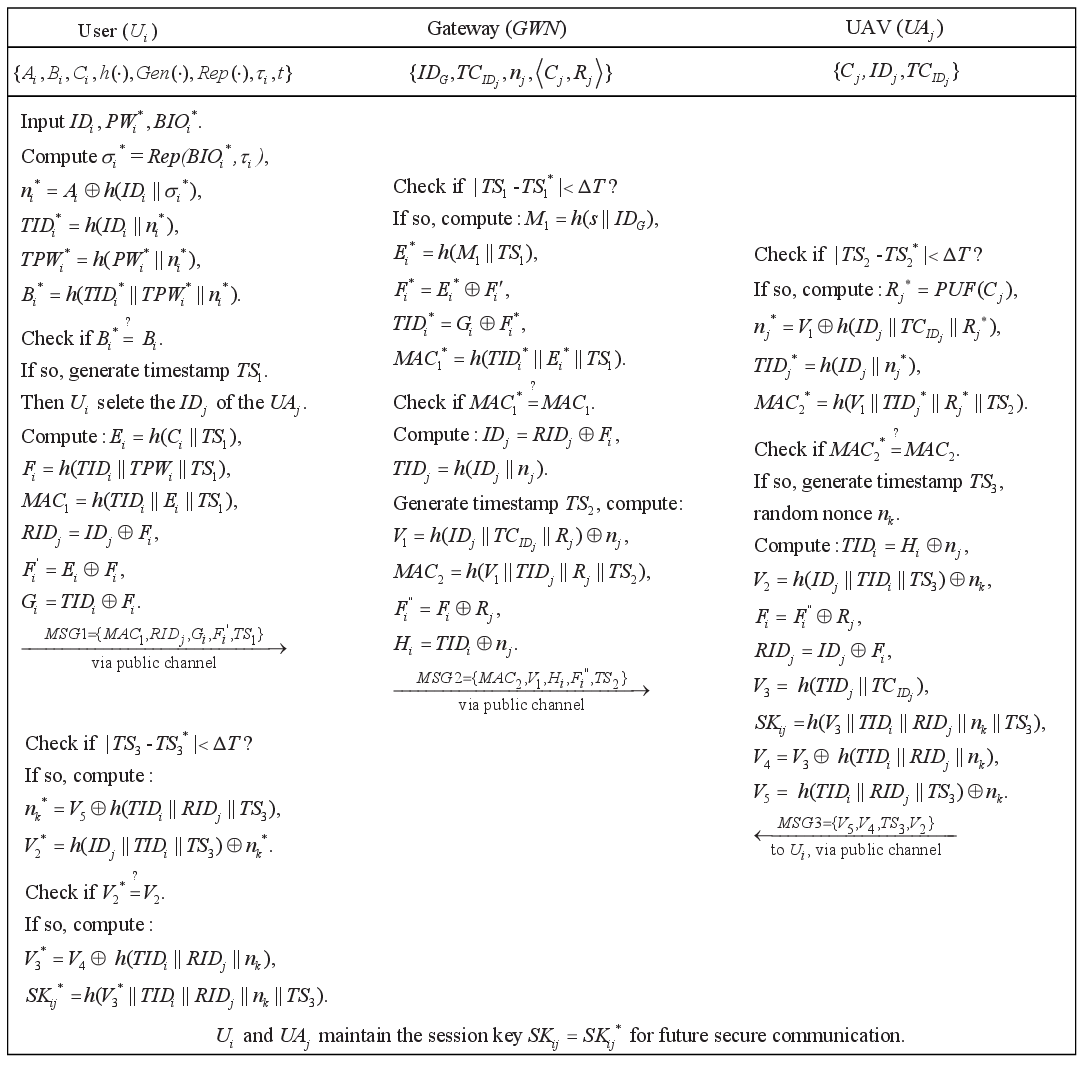}
\captionsetup{font={small},labelsep=period}
\caption{\label{AKA}\raggedright Login and AKA phases.}
\end{figure}

\subsection {Authentication and Key Agreement}

The AKA phase in the proposed protocol establishes a secure session key \( SK_{ij} \) among the  user \( U_i \),  GWN, and UAV  \( UA_j \). The specific login and authentication process is shown in Fig. \ref{AKA}.

\subsubsection{Step 1}  

User \( U_i \) begins authentication by inputting their identity \( ID_i \), password \( PW_i^* \), and biometric data \( BIO_i^* \). From \( BIO_i^* \), \( U_i \) derives the biometric-based secret \( \sigma_i^* = Rep(BIO_i^*, \tau_i) \), which is used to compute the temporary secret \( n_i^* = A_i \oplus h(ID_i || \sigma_i^*) \). Then, \( U_i \) calculates the temporary identity \( TID_i^* = h(ID_i || n_i^*) \), the transformed password \( TPW_i^* = h(PW_i^* || n_i^*) \), and the authentication verifier \( B_i^* = h(TID_i^* || TPW_i^* || n_i^*) \), verifying \( B_i^* \stackrel{?}{=} B_i \). After selecting target UAV \( UA_j \) identified by \( ID_j \), \( U_i \) generates timestamp \( TS_1 \) and computes intermediate values: \( E_i = h(C_i || TS_1) \), \( F_i = h(TID_i || TPW_i || TS_1) \), \( MAC_1 = h(TID_i || E_i || TS_1) \), \( RID_j = ID_j \oplus F_i \), \( F_i^\prime = E_i \oplus F_i \), and \( G_i = TID_i \oplus F_i \). These are sent as message \( MSG1 = \{ MAC_1, RID_j, G_i, F_i^\prime, TS_1 \} \) to the GWN via a public channel.

\subsubsection{Step 2}  

Upon receiving \( MSG1 \), the GWN verifies the freshness of \( TS_1 \) by checking \( |TS_1 - TS_1^*| < \Delta T \). If valid, it computes \( M_1 = h(s || ID_G) \) and derives \( E_i^* = h(M_1 || TS_1) \), \( F_i^* = E_i^* \oplus F_i^\prime \), and \( TID_i^* = G_i \oplus F_i^* \). The GWN confirms authenticity by checking \( MAC_1^* = h(TID_i^* || E_i^* || TS_1) \stackrel{?}{=} MAC_1 \). If successful, it retrieves \( ID_j = RID_j \oplus F_i \) and computes \( TID_j = h(ID_j || n_j) \). The GWN generates a new timestamp \( TS_2 \) and computes: \( V_1 = h(ID_j || TC_{ID_j} || R_j) \oplus n_j \), \( MAC_2 = h(V_1 || TID_j || R_j || TS_2) \), \( F_i^{\prime\prime} = F_i \oplus R_j \), and \( H_i = TID_i \oplus n_j \). The message \( MSG2 = \{ MAC_2, V_1, H_i, F_i^{\prime\prime}, TS_2 \} \) is sent to UAV \( UA_j \) via a public channel.

\subsubsection{Step 3}  

Upon receiving \( MSG2 \), UAV \( UA_j \) checks the freshness of \( TS_2 \) by verifying \( |TS_2 - TS_2^*| < \Delta T \). Using its PUF, it computes \( R_j^* = PUF(C_j) \), derives \( n_j^* = V_1 \oplus h(ID_j || TC_{ID_j} || R_j^*) \), and calculates \( TID_j^* = h(ID_j || n_j^*) \). UAV \( UA_j \) checks the integrity of the message by verifying \( MAC_2^* = h(V_1 || TID_j^* || R_j^* || TS_2) \stackrel{?}{=} MAC_2 \). It generates a random nonce \( n_k \), a new timestamp \( TS_3 \), and computes: \( TID_i = H_i \oplus n_j \), \( V_2 = h(ID_j || TID_i || TS_3) \oplus n_k \), \( F_i = F_i^{\prime\prime} \oplus R_j \), \( RID_j = ID_j \oplus F_i \), \( V_3 = h(TID_j || TC_{ID_j}) \), \( SK_{ij} = h(V_3 || TID_i || RID_j || n_k || TS_3) \), \( V_4 = V_3 \oplus h(TID_i || RID_j || n_k) \), and \( V_5 = h(TID_i || RID_j || TS_3) \oplus n_k \). UAV \( UA_j \) sends \( MSG3 = \{ V_5, V_4, TS_3, V_2 \} \) to user \( U_i \) via a public channel.

\subsubsection{Step 4}  

User \( U_i \) processes \( MSG3 \) to finalize the session key \( SK_{ij} \). It validates the freshness of \( TS_3 \) by checking \( |TS_3 - TS_3^*| < \Delta T \), computes \( n_k^* = V_5 \oplus h(TID_i || RID_j || TS_3) \), derives \( V_2^* = h(ID_j || TID_i || TS_3) \oplus n_k^* \), and verifies \( V_2^* \stackrel{?}{=} V_2 \). If the check passes, it calculates \( V_3^* = V_4 \oplus h(TID_i || RID_j || n_k) \) and establishes the session key \( SK_{ij}^* = h(V_3^* || TID_i || RID_j || n_k || TS_3) \). Since \( SK_{ij}^* = SK_{ij} \), the session key is successfully shared between user \( U_i \) and UAV \( UA_j \), concluding the mutual authentication process and enabling secure future communications.

\subsection{Password and Biometric Update}

\subsubsection{Step 1}
The user inserts the smart card, providing their identifier \( ID_i \), old password \( PW_i^{\text{old}} \), and old biometric data \( BIO_i^{\text{old}} \). The system computes the old biometric key \( \sigma_i^{\text{old}} = Rep(BIO_i^{\text{old}}, \tau_i) \), and calculates \( n_i = A_i \oplus h(ID_i \| \sigma_i^{\text{old}}) \), \( TID_i = h(ID_i \| n_i) \), and \( TPW_i^{\text{old}} = h(PW_i^{\text{old}} \| n_i) \). The system verifies the user by checking if \( B_i^{\text{old}} \stackrel{?}{=} B_i \), where \( B_i^{\text{old}} = h(ID_i \| TPW_i^{\text{old}} \| \sigma_i^{\text{old}}) \). If successful, the user is authenticated and can proceed to update their credentials.

\subsubsection{Step 2}
After authentication, the user provides a new password \( PW_i^{\text{new}} \) and updated biometric data \( BIO_i^{\text{new}} \). The system computes the new biometric key and reproduction parameter \( (\sigma_i^{\text{new}}, \tau_i^{\text{new}}) = Gen(BIO_i^{\text{new}}) \), and updates the following values: \( TPW_i^{\text{new}} = h(PW_i^{\text{new}} \| n_i) \), \( A_i^{\text{new}} = n_i \oplus h(ID_i \| \sigma_i^{\text{new}}) \), \( B_i^{\text{new}} = h(ID_i \| TPW_i^{\text{new}} \| \sigma_i^{\text{new}}) \), and \( C_i^{\text{new}} = TC_{ID_i} \oplus TID_i \oplus TPW_i^{\text{new}} \). These updated values are used for subsequent operations.

\subsubsection{Step 3}
Finally, the user updates the smart card \( SC_i \) by replacing the old parameters with the new ones: \( A_i = A_i^{\text{new}} \), \( B_i = B_i^{\text{new}} \), \( C_i = C_i^{\text{new}} \), and \( \tau_i = \tau_i^{\text{new}} \).

\subsection{Smart Card Replacement}

\subsubsection{Step 1}
User \( U_i \) enters their identifier \( ID_i \) and a new password \( PW_i^{\prime} \), selects a random nonce \( n_i^{\prime} \), and computes \( TID_i = h(ID_i \| n_i^{\prime}) \) and \( TPW_i^{\prime} = h(PW_i^{\prime} \| n_i^{\prime}) \). The user then submits a revocation request containing \( \{TID_i, TPW_i^{\prime}\} \) to the GWN via a secure channel.

\subsubsection{Step 2}
Upon receiving the request, the GWN verifies whether \( TID_i \) exists in its database. If no match is found, the gateway allocates a new smart card \( SC_i^{\text{new}} \) and sends it for activation.

\subsubsection{Step 3}
Once user \( U_i \) receives the new smart card, they input their biometric key \( BIO_i \) and compute the new \( \sigma_i \) and \( \tau_i \). The user then calculates the new authentication values: \( A_i = n_i^{\prime} \oplus h(ID_i \| \sigma_i) \), \( B_i = h(ID_i \| TPW_i^{\prime} \| \sigma_i) \), and \( C_i = TC_{ID_i} \oplus TID_i \oplus TPW_i^{\prime} \). These values are stored in the new smart card, which is updated as \( SC_i^{\text{new}} = \{A_i, B_i, C_i, h(\cdot), Gen(\cdot), Rep(\cdot), \tau_i \} \).

\subsection{Dynamic UAV Addition}

\subsubsection{Step 1}
The GWN assigns a unique identifier \( ID_j^{new} \) to the newly added UAV. The UAV initiates the registration process by securely sending \( ID_j^{new} \) to the gateway via a protected communication channel. The gateway generates a random nonce \( n_j^{new} \) and computes the temporary identifier \( TID_j^{new} = h(ID_j^{new} \| n_j^{new}) \). It then creates the temporary certificate \( TC_{ID_j^{new}} = h(TID_j^{new} \| s) \) and a challenge \( C_j^{new} \), which are securely sent to the UAV to complete the registration.

\subsubsection{Step 2}
Upon receiving the challenge \( C_j^{new} \), the UAV uses its PUF to compute the response \( R_j^{new} = PUF(C_j^{new}) \). The UAV sends the response back to the gateway and stores \( C_j^{new} \), \( ID_j^{new} \), and \( TC_{ID_j^{new}} \). The gateway similarly stores \( TC_{ID_j^{new}} \), \( n_j^{new} \), and the challenge-response pair \( \langle C_j^{new}, R_j^{new} \rangle \) in its secure database. After registration, the gateway broadcasts the new UAV information to all registered users, allowing them to access the UAV node based on their needs, ensuring smooth integration into the system.

\section{Security Analysis}
\label{sec5}

\subsubsection{Smart Card Theft Attack}
In this scenario, assume that the adversary \( \mathcal{A} \) gains access to the smart card of a registered user \( U_i \). Through power analysis, the adversary can extract information \( \{ A_i, B_i, C_i, h(\cdot), \text{Gen}(\cdot), \text{Rep}(\cdot), \tau_i, t \} \) from the card. However, without the random nonce \( n_i \), it is computationally infeasible for \( \mathcal{A} \) to derive the identity \( ID_i \) or password \( PW_i \) of the user, ensuring user anonymity.

\subsubsection{Privileged Insider Attack}
Assume that \( \mathcal{A} \) is a malicious insider with trusted access to the GWN. During registration, \( U_i \) securely transmits the request \( \{ TID_i, TPW_i \} \) to the GWN. Even if \( \mathcal{A} \) intercepts this message, the collision-resistant properties of the hash function \( h(\cdot) \) make it computationally infeasible to derive \( PW_i \) without knowledge of the random key \( n_i \) and \( ID_i \), preventing a successful attack.

\subsubsection{Impersonation Attacks}
In impersonation attacks, \( \mathcal{A} \) attempts to impersonate a legitimate entity by intercepting and forging authentication messages. However, valid message construction requires knowledge of confidential parameters specific to each entity. For example, impersonating the user requires computing \( F_i' \) and \( G_i \), which depend on private parameters unknown to \( \mathcal{A} \). Similarly, impersonating the GWN requires access to \( n_j \) and \( R_j \) to generate a valid \( MSG2 \), and impersonating a UAV requires random keys \( n_j \) and \( n_k \) to construct \( MSG3 \). Since these parameters remain inaccessible, impersonation attacks are thwarted.

\subsubsection{User Anonymity and Untraceability}
Even if adversary \( \mathcal{A} \) intercepts communication messages \( MSG1 \), \( MSG2 \), and \( MSG3 \), deriving the true identity \( ID_i \) of the user is computationally infeasible. The messages rely on secret parameters such as \( F_i \), \( R_j \), and private keys \( n_j \) and \( n_k \), which are inaccessible to \( \mathcal{A} \), ensuring user anonymity and untraceability.

\subsubsection{Resilience Against UAV Capture Attack}
In the event that \( \mathcal{A} \) captures a UAV \( UA_j \), it may extract credentials stored in the UAV’s memory, such as \( \{C_j, ID_j, TC_{ID_j}\} \). However, cryptographic hashing ensures that sensitive parameters like \( TID_j \) and \( TC_{ID_j} \) are computationally infeasible to reverse-engineer without the private key \( s \) and random secret \( n_j \), maintaining the security of session keys and communication between non-compromised UAVs and users.

\subsubsection{Mutual Authentication}
The protocol ensures mutual authentication among \( U_i \), GWN, and \( UA_j \) through a three-step process. 
Firstly, \( U_i \) generates temporary credentials: hashed identity \( TID_i^* \), temporary password \( TPW_i^* \), and an authentication code \( MAC_1 \), which are sent to GWN for verification.
Then, GWN authenticates itself to \( UA_j \) by sending \( MSG2 \), containing the verification code \( MAC_2 \), encrypted parameters, and a timestamp. 
 Finally, \( UA_j \) authenticates itself to \( U_i \) by sending \( MSG3 \), containing a session key component and \( MAC_3 \). Upon verification, \( U_i \) confirms the authenticity of \( UA_j \). A secure session key \( SK_{ij} \) is established.

\subsubsection{Replay Attack}
The protocol mitigates replay attacks by using time-sensitive parameters and one-time authentication tokens in each communication stage. Each message, such as \( MSG1 \), \( MSG2 \), and \( MSG3 \), contains a timestamp to ensure data freshness. The GWN and \( UA_j \) verify timestamps against a predefined threshold \( \Delta T \), preventing replayed messages from being reused by an adversary.

\subsubsection{Man-in-the-Middle (MITM) Attack}
The protocol prevents MITM attacks through mutual authentication. Each message exchange between \( U_i \), GWN, and \( UA_j \) includes authentication codes, ensuring that any tampered message is rejected. This restricts communication to authenticated parties, protecting against MITM attacks.

\subsubsection{Ephemeral Secret Leakage Attack}
The protocol protects against ephemeral secret leakage by securely combining short-term and long-term secrets in the session key derivation process. Even if adversaries gain access to short-term secrets, such as \( n_j \) or \( n_k \), they cannot compute the session key \( SK_{ij} \) without additional long-term credentials. Dynamic, time-sensitive parameters ensure forward and backward security, preventing the compromise of one session key from affecting others.

\subsubsection{Denial of Service (DoS) Attack}
The protocol mitigates DoS attacks by validating timestamps and message authentication codes at each stage of authentication. Invalid timestamps or mismatched authentication codes cause the protocol to terminate the process, preventing malicious entities from overwhelming the system with invalid requests.

\subsubsection{Side Channel Attack}
The protocol defends against side-channel attacks through the use of PUFs. During authentication, the UAV generates a response \( R_j^* = PUF(C_j) \), which is resistant to side-channel attacks like power analysis or electromagnetic emissions. The uniqueness of the PUF for each UAV ensures that adversaries cannot derive meaningful information from side-channel observations.

\subsubsection{Challenge-Response Pair (CRP) Leakage Attack}
The protocol mitigates CRP leakage by dynamically generating responses during authentication and cryptographically binding them to session parameters. Since the UAV stores only the challenge \( C \) and computes the response \( R_j^* \) at authentication time, adversaries cannot access the responses directly. The inherent unpredictability of the PUF further protects against CRP leakage attacks.

\begin{table*}[t]
\centering
\renewcommand{\arraystretch}{1.5}
\captionsetup{justification=centering, labelformat=default,labelsep=newline,textfont=sc}
\caption{Comparison of Computation Overheads.}
\resizebox{\textwidth}{!}{
\begin{tabular}{@{}cllll@{}}
\toprule
\textbf{Protocol} & \textbf{User side} & \textbf{GWN/ Server side} & \textbf{Sensor/ UAV side} & \textbf{Total cost} \\ \midrule
\textbf{Ref \cite{wazid2019design}}      & $1T_{fe} + 16T_h \approx 0.648ms$ & $8T_h \approx 0.008ms$ & $7T_h \approx 0.007ms$ & $1T_{fe}  + 31T_h \approx 0.663ms$ \\
\textbf{Ref \cite{xu2023three}}     & $2T_{eca} + 5T_{ecm} + 6T_h \approx 3.198ms$ & $1T_{eca} + 1T_{ecm} + 4T_h \approx 0.652ms$ & $5T_{eca} + 7T_{ecm} + 5T_h \approx 4.509ms$ & $8T_{eca} + 13T_{ecm} + 15T_h \approx 8.359ms$ \\
\textbf{\makecell[c]{Ref \cite{alladi2020harci}}} & 
\makecell[l]{$1T_{enc} + 6T_{bp} + 2T_{hmac}$\\
$ + 1T_{puf} + 2T_h \approx 26.049ms$} & 
\makecell[l]{$3T_{enc} + 9T_{bp} + 3T_{hmac}$\\
$ + 1T_{puf} + 2T_h \approx 39.14ms$} & 
\makecell[l]{$7T_{enc} + 6T_{bp} + 3T_{hmac}$\\
$ + 2T_h \approx 26.422ms$} & 
\makecell[l]{$11T_{enc} + 21T_{bp} + 8T_{hmac}$\\
$ + 2T_{puf} + 6T_h \approx 91.611ms$} \\
\textbf{Ref \cite{mahmood2022provably}}     & $4T_h \approx 0.004ms$ & $1T_{ecm} + 5T_h \approx 0.637ms$ & $1T_{puf} + 1T_{ecm} + 4T_h \approx 0.651ms$ & $2T_{ecm} + 1T_{puf} + 13T_h \approx 1.292ms$  \\ 
\textbf{Proposed}      & $1T_{fe} + 11T_h \approx 0.643ms$ & $6T_h \approx 0.006ms$ & $1T_{puf} + 8T_h \approx 0.023ms$ & $1T_{fe} + 1T_{puf} + 25T_h \approx 0.672ms$ \\
\bottomrule
\end{tabular}
}

\label{tab:computation-overheads}
\end{table*}

\section{Performance Comparison}
\label{sec6}
We compare the proposed protocol with other related works in the aspects of computing
and communication overhead in this section. Besides, we also analyzed the safety performance.

\begin{table}[t]
\centering
\captionsetup{justification=centering, labelformat=default,labelsep=newline,textfont=sc}
\caption{Comparison of Communication Overheads.}
\label{tab:communication-overheads}
\begin{tabular*}{0.9\columnwidth}{@{\extracolsep{\fill}} ccc @{}} 
\toprule
\textbf{Protocol} & \textbf{No. of messages} & \textbf{No. of bits} \\ \midrule
\textbf{\makecell[c]{Ref \cite{wazid2019design}}} & \makecell[c]{3} & \makecell[c]{1696} \\
\textbf{\makecell[c]{Ref \cite{xu2023three}}}     & \makecell[c]{3} & \makecell[c]{2336} \\
\textbf{\makecell[c]{Ref \cite{alladi2020harci}}} & \makecell[c]{6} & \makecell[c]{3200} \\
\textbf{\makecell[c]{Ref \cite{mahmood2022provably}}} & \makecell[c]{3} & \makecell[c]{2240} \\
\textbf{\makecell[c]{Proposed}}                         & \makecell[c]{3} & \makecell[c]{1856} \\ \bottomrule
\end{tabular*}
\end{table}

\subsection{Computational Overhead}  

Computational overhead is evaluated based on the time consumption of cryptographic operations. Here, $ T_{ecm} $ and $ T_{eca} $ denote multiplication and addition operations on the cyclic group, while $ T_{fe} $, $ T_h $, $ T_{puf} $, $ T_{enc} $, $ T_{bp} $, and $ T_{hmac} $ correspond to the fuzzy extractor function, hash function, PUF operation, symmetric encryption, bilinear pairing, and hash-based message authentication, respectively. The XOR operation is negligible.

Simulation experiments using the Java pairing-based cryptography (JPBC) library \cite{de2011jpbc} on a laptop (16 GB RAM, 3.1 GHz CPU) yielded the following time consumptions: \( T_{ecm} = 0.632 \)ms, \( T_{eca} = 0.016 \)ms, \( T_h = 0.001 \)ms, \( T_{puf} = 0.015 \)ms, \( T_{enc} = 0.05 \)ms, \( T_{bp} = 4.301 \)ms, and \( T_{hmac} = 0.088 \)ms \cite{wang2024authentication}. We adopt \( T_{fe} \approx T_{ecm} \) as in \cite{wazid2019design}.

Table \ref{tab:computation-overheads} compares computational overheads across user, gateway/server, and UAV/sensor components. On the user side, our protocol requires 0.643ms (1\( T_{fe} \) and 11\( T_h \)), comparable to \cite{wazid2019design} (0.648ms) and significantly lower than \cite{xu2023three} (3.198 ms) and \cite{alladi2020harci} (26.049ms), which involve bilinear pairing and symmetric encryption. On the gateway/server side, our protocol incurs only 0.006ms, outperforming \cite{xu2023three} (0.652ms), \cite{alladi2020harci} (39.14ms), and \cite{mahmood2022provably} (0.637ms). For UAVs/sensors, our protocol requires 0.023ms, lower than 0.651ms in \cite{mahmood2022provably} but higher than 0.007ms in \cite{wazid2019design} due to the PUF operation enhancing security.

Overall, the proposed protocol achieves low computational overhead while maintaining robust security.

\subsection{Communication Overhead}  

Communication overhead is critical in authentication protocols, affecting bandwidth utilization and transmission delays. Identity, random nonce, hash output (SHA-1), and timestamp are considered as 160, 128, 160, and 32 bits, respectively.

Table \ref{tab:communication-overheads} compares the communication overhead of various protocols, considering both message count and bit size. Our protocol exchanges three messages: \( \text{MSG1} = \{ \text{MAC}_1, \text{RID}_j, G_i, F_i^\prime, \text{TS}_1 \} \) from \( U_i \) to GWN (672 bits), \( \text{MSG2} = \{ \text{MAC}_2, V_1, H_i, F_i^{\prime\prime}, \text{TS}_2 \} \) from GWN to \( UA_j \) (672 bits), and \( \text{MSG3} = \{ V_5, V_4, \text{TS}_3, V_2 \} \) from \( UA_j \) to \( U_i \) (512 bits), totaling 1,856 bits.

Compared to \cite{wazid2019design} (1,696 bits), our protocol provides enhanced security with comparable efficiency. Protocols in \cite{xu2023three}, \cite{alladi2020harci}, and \cite{mahmood2022provably} incur higher overheads of 2,336, 3,200, and 2,240 bits, respectively, demonstrating the superior efficiency of our approach.

\subsection{Security and Functionality Features}  

Informal security analysis confirms our protocol defends against common attacks while supporting key features such as password and biometric key updates, smartcard revocation, and dynamic UAV deployment. Table \ref{tab:comparison} highlights its advantages, including protection against ephemeral secret leakage, which is unaddressed by other protocols. Unlike \cite{alladi2020harci} and \cite{mahmood2022provably}, which lack support for dynamic sensor addition, our protocol seamlessly integrates new devices.

Compared to existing solutions, our protocol achieves superior security and functionality with minimal computational and communication overhead.

\begin{table}[t]
    \centering
    \renewcommand{\arraystretch}{1.2}
    \captionsetup{justification=centering, labelformat=default,labelsep=newline,textfont=sc}
    \caption{Comparison of Functionality and Security Features.}
    \label{tab:comparison}
    
    \begin{tabular*}{\columnwidth}{@{\extracolsep{\fill}}lccccc} 
        \hline
        \textbf{Feature} & \textbf{Ref \cite{wazid2019design}} & \textbf{Ref \cite{xu2023three}} & \textbf{Ref \cite{alladi2020harci}} & \textbf{Ref \cite{mahmood2022provably}} & \textbf{Proposed} \\
        \hline
        $FSF_1$          & $\checkmark$                    & $\checkmark$               & $\times$  & $\checkmark$  & $\checkmark$  \\
        $FSF_2$          & $\checkmark$                        & $\checkmark$               & $\times$  & $\times$  & $\checkmark$  \\
        $FSF_3$          & $\checkmark$                        & $\checkmark$               & $\checkmark$  & $\checkmark$  & $\checkmark$  \\
        $FSF_4$          & $\checkmark$                        & $\checkmark$               & $\checkmark$  & $\checkmark$  & $\checkmark$  \\
        $FSF_5$          & $\checkmark$                    & $\checkmark$               & $\checkmark$  & $\checkmark$  & $\checkmark$  \\
        $FSF_6$          & $\checkmark$                    & $\checkmark$               & $\checkmark$  & $\checkmark$  & $\checkmark$  \\
        $FSF_7$          & $\checkmark$                    & $\checkmark$               & $\checkmark$  & $\checkmark$  & $\checkmark$  \\
        $FSF_{8}$       & $\checkmark$                    & $\checkmark$               & $\checkmark$  & $\checkmark$  & $\checkmark$  \\
        $FSF_{9}$       & $\checkmark$                        & $\checkmark$               & $\times$  & $\times$  & $\checkmark$  \\
        $FSF_{10}$       & $\checkmark$                    & $\checkmark$               & $\times$  & $\times$  & $\checkmark$  \\
        $FSF_{11}$       & $\times$                    & $\times$               & $\checkmark$  & $\checkmark$  & $\checkmark$  \\
        $FSF_{12}$       & $\times$                        & $\times$               & $\times$  & $\times$  & $\checkmark$  \\
        $FSF_{13}$       & $\checkmark$                        & $\times$               & $\times$  & $\times$  & $\checkmark$  \\
        $FSF_{14}$       & $\checkmark$                        & $\checkmark$                   & $\times$  & $\times$  & $\checkmark$  \\
     
        \hline
    \end{tabular*}

    \smallskip
\noindent
\begin{tabular}{@{}p{\linewidth}@{}}
\textbf{Note:} 
$FSF_1$: Stolen Mobile/Smart Card Attack; $FSF_2$: Privileged-Insider attack; $FSF_3$: Impersonation Attack;  $FSF_4$: User Anonymity and Untraceability; $FSF_5$: Sensing Node Capture Attack; $FSF_6$: Mutual Authentication; $FSF_7$: Replay Attack; $FSF_{8}$:  MITM Attack; $FSF_{9}$: Ephemeral Secret Leakage  Attack; $FSF_{10}$: DoS Attack; $FSF_{11}$: Side Channel Attack; $FSF_{12}$: CRP Leakage Attack; $FSF_{13}$: Support Password and Biological Key Update; $FSF_{14}$:  Supports Adding Sensor Devices; \\
$\times$: Insecure against a particular attack or does not support a particular feature; \\
$\checkmark$: Secure against a particular attack or supports a particular feature.
\end{tabular}
    
\end{table}

\section{Conclusion}
\label{sec8}

In this work, we have proposed a lightweight PUF-based AKA protocol tailored for FANET. By leveraging the unique properties of PUFs, the protocol has provided secure and efficient authentication, addressing the challenges posed by the dynamic and resource-constrained nature of FANET environments. Through  security analysis, we have demonstrated the resilience of the protocol against various potential attacks. Comparative evaluations have shown that the proposed protocol significantly outperforms existing protocols in terms of both the security and efficiency, reducing computational and communication overheads.

\small
\bibliographystyle{IEEEtranN}

\bibliography{wu2024adaptive,wazid2019design,xu2023three,alladi2020harci,mahmood2022provably,de2011jpbc,wang2024authentication,
odelu2015efficient,dolev1983security,won2015secure,messerges2002examining,lamport1981password,das2009two,he2010enhanced,
alladi2020parth,srinivas2019tcalas,ali2020securing,won2017certificateless,cheon2018toward,ever2020secure,nikooghadam2021provably,
zhang2020lightweight,gope2020efficient,abdalla2005password,riley2010ns,chang2015provably,zhou2023aerospace,lu2024joint,jia2024cooperative,
he2023routing,zhang2024prlap,yaacoub2020security,liang2021mutual,chen2022ecc,wang2023secure,messaoudi2023survey,yang2023depth,
pattaranantakul2023service,du2024tri,jia2025hierarchical,11122503,jia2025distributionally}

\begin{thebibliography}{20}
\providecommand{\natexlab}[1]{#1}
\providecommand{\url}[1]{#1}
\csname url@samestyle\endcsname
\providecommand{\newblock}{\relax}
\providecommand{\bibinfo}[2]{#2}
\providecommand{\BIBentrySTDinterwordspacing}{\spaceskip=0pt\relax}
\providecommand{\BIBentryALTinterwordstretchfactor}{4}
\providecommand{\BIBentryALTinterwordspacing}{\spaceskip=\fontdimen2\font plus
\BIBentryALTinterwordstretchfactor\fontdimen3\font minus
  \fontdimen4\font\relax}
\providecommand{\BIBforeignlanguage}[2]{{%
\expandafter\ifx\csname l@#1\endcsname\relax
\typeout{** WARNING: IEEEtranN.bst: No hyphenation pattern has been}%
\typeout{** loaded for the language `#1'. Using the pattern for}%
\typeout{** the default language instead.}%
\else
\language=\csname l@#1\endcsname
\fi
#2}}
\providecommand{\BIBdecl}{\relax}
\BIBdecl

\bibitem[Yang et~al.(Jan. 2024)Yang, Hu, Yu, Chen, and Xu]{yang2023depth}
J.~Yang, Y.~Hu, Z.~Yu, F.~Chen, and X.~Xu, ``In-depth coordination and
  extension: Decentralized onboard conflict resolution of {UAVs} in the low
  altitude airspace,'' \emph{IEEE Trans. Intell. Veh.}, vol.~9, no.~1, pp.
  2780--2793, Jan. 2024.

\bibitem[Jia et~al.(May 2025)Jia, Cui, Dong, Wu, Ling, Niyato, and
  Han]{jia2025distributionally}
Z.~Jia, C.~Cui, C.~Dong, Q.~Wu, Z.~Ling, D.~Niyato, and Z.~Han,
  ``Distributionally robust optimization for aerial multi-access edge computing
  via cooperation of {UAVs and HAPs},'' \emph{IEEE Trans. Mob. Comput.},
  vol.~24, no.~10, pp. 10\,853--10\,867, May 2025.

\bibitem[Zhou et~al.(Feb. 2023)Zhou, Sheng, Li, and Han]{zhou2023aerospace}
D.~Zhou, M.~Sheng, J.~Li, and Z.~Han, ``Aerospace integrated networks
  innovation for empowering {6G}: A survey and future challenges,'' \emph{IEEE
  Commun. Surveys Tut.}, vol.~25, no.~2, pp. 975--1019, Feb. 2023.

\bibitem[Lu et~al.(Oct. 2024)Lu, Jia, Wu, and Han]{lu2024joint}
Z.~Lu, Z.~Jia, Q.~Wu, and Z.~Han, ``Joint trajectory planning and communication
  design for multiple {UAVs} in intelligent collaborative air-ground
  communication systems,'' \emph{IEEE Internet Things J.}, vol.~11, no.~19, pp.
  31\,053--31\,067, Oct. 2024.

\bibitem[Jia et~al.(Aug. 2025)Jia, He, Zhu, Wang, Wu, and Han]{11122503}
Z.~Jia, S.~He, Q.~Zhu, W.~Wang, Q.~Wu, and Z.~Han, ``Trusted routing for
  blockchain-empowered {UAV} networks via multi-agent deep reinforcement
  learning,'' \emph{IEEE Trans. Commun.}, Aug. 2025, early access.

\bibitem[Pattaranantakul et~al.(Feb. 2023)Pattaranantakul, Vorakulpipat, and
  Takahashi]{pattaranantakul2023service}
M.~Pattaranantakul, C.~Vorakulpipat, and T.~Takahashi, ``Service function
  chaining security survey: Addressing security challenges and threats,''
  \emph{Comput. Networks}, vol. 221, p. 109484, Feb. 2023.

\bibitem[Wu et~al.(Aug. 2024)Wu, Jia, Wu, and Lu]{wu2024adaptive}
Y.~Wu, Z.~Jia, Q.~Wu, and Z.~Lu, ``Adaptive {QoE}-aware {SFC} orchestration in
  {UAV} networks: A deep reinforcement learning approach,'' \emph{IEEE Trans.
  Network Sci. Eng.}, vol.~11, no.~6, pp. 6052--6065, Aug. 2024.

\bibitem[Du et~al.(Mar. 2024)Du, Cao, Wang, Lv, Wu, and Wang]{du2024tri}
X.~Du, Y.~Cao, D.~Wang, C.~Lv, C.~Wu, and K.~Wang, ``A {Tri-Phases} message
  oriented trust model in {FANET},'' \emph{IEEE Trans. Network Sci. Eng.},
  vol.~11, no.~6, pp. 5298--5310, Mar. 2024.

\bibitem[Yaacoub et~al.(Sep. 2020)Yaacoub, Noura, Salman, and
  Chehab]{yaacoub2020security}
J.-P. Yaacoub, H.~Noura, O.~Salman, and A.~Chehab, ``Security analysis of
  drones systems: Attacks, limitations, and recommendations,'' \emph{Internet
  Things}, vol.~11, p. 100218, Sep. 2020.

\bibitem[Jia et~al.(Sep. 2024)Jia, You, Dong, Wu, Zhou, Niyato, and
  Han]{jia2024cooperative}
Z.~Jia, J.~You, C.~Dong, Q.~Wu, F.~Zhou, D.~Niyato, and Z.~Han, ``Cooperative
  cognitive dynamic system in {UAV} swarms: Reconfigurable mechanism and
  framework,'' \emph{IEEE Veh. Technol. Mag.}, vol.~19, no.~3, pp. 90--101,
  Sep. 2024.

\bibitem[Liang et~al.(Oct. 2021)Liang, Xie, Zhang, Li, and Li]{liang2021mutual}
W.~Liang, S.~Xie, D.~Zhang, X.~Li, and K.-C. Li, ``A mutual security
  authentication method for {RFID-PUF} circuit based on deep learning,''
  \emph{ACM Trans. Internet Technol.}, vol.~22, no.~2, pp. 1--20, Oct. 2021.

\bibitem[Dolev and Yao(Mar. 1983)]{dolev1983security}
D.~Dolev and A.~Yao, ``On the security of public key protocols,'' \emph{IEEE
  Trans. Inf. Theory}, vol.~29, no.~2, pp. 198--208, Mar. 1983.

\bibitem[Jia et~al.(2025)Jia, He, Cui, Zhu, Yuan, Zhou, Wu, Niyato, and
  Han]{jia2025hierarchical}
Z.~Jia, J.~He, Y.~Cui, Q.~Zhu, L.~Yuan, F.~Zhou, Q.~Wu, D.~Niyato, and Z.~Han,
  ``Hierarchical low-altitude wireless network empowered air traffic
  management,'' \emph{arXiv preprint arXiv:2509.03386}, 2025.

\bibitem[Messerges et~al.(May. 2002)Messerges, Dabbish, and
  Sloan]{messerges2002examining}
T.~S. Messerges, E.~A. Dabbish, and R.~H. Sloan, ``Examining smart-card
  security under the threat of power analysis attacks,'' \emph{IEEE Trans.
  Comput.}, vol.~51, no.~5, pp. 541--552, May. 2002.

\bibitem[Wazid et~al.(Apr. 2019)Wazid, Das, Kumar, Vasilakos, and
  Rodrigues]{wazid2019design}
M.~Wazid, A.~K. Das, N.~Kumar, A.~V. Vasilakos, and J.~J. P.~C. Rodrigues,
  ``Design and analysis of secure lightweight remote user authentication and
  key agreement scheme in {Internet of Drones} deployment,'' \emph{IEEE
  Internet Things J.}, vol.~6, no.~2, pp. 3572--3584, Apr. 2019.

\bibitem[Xu et~al.(Aug. 2023)Xu, Hsu, Harn, Cui, Zhao, and Zhang]{xu2023three}
H.~Xu, C.~Hsu, L.~Harn, J.~Cui, Z.~Zhao, and Z.~Zhang, ``Three-factor anonymous
  authentication and key agreement based on fuzzy biological extraction for
  {Industrial Internet of Things},'' \emph{IEEE Trans. Services Comput.},
  vol.~16, no.~4, pp. 3000--3013, Aug. 2023.

\bibitem[Alladi et~al.(Feb. 2021)Alladi, Chamola, and Naren]{alladi2020harci}
T.~Alladi, V.~Chamola, and Naren, ``{HARCI}: A two-way authentication protocol
  for three entity healthcare {IoT} networks,'' \emph{IEEE J. Sel. Areas
  Commun.}, vol.~39, no.~2, pp. 361--369, Feb. 2021.

\bibitem[Mahmood et~al.(Feb. 2023)Mahmood, Ferzund, Saleem, Shamshad, Das, and
  Park]{mahmood2022provably}
K.~Mahmood, J.~Ferzund, M.~A. Saleem, S.~Shamshad, A.~K. Das, and Y.~Park, ``A
  provably secure mobile user authentication scheme for big data collection in
  {IoT-enabled} maritime intelligent transportation system,'' \emph{IEEE Trans.
  Intell. Transp. Syst.}, vol.~24, no.~2, pp. 2411--2421, Feb. 2023.

\bibitem[De~Caro and Iovino(2011)]{de2011jpbc}
A.~De~Caro and V.~Iovino, ``{JPBC: Java} pairing based cryptography,'' in
  \emph{Proc. IEEE Symp. Comput. Commun. (ISCC)}, Kerkyra, Greece, Aug. 2011,
  pp. 850--855.

\bibitem[Wang et~al.(Jun. 2024)Wang, Cao, Lam, Hu, and
  Kaiwartya]{wang2024authentication}
D.~Wang, Y.~Cao, K.-Y. Lam, Y.~Hu, and O.~Kaiwartya, ``Authentication and key
  agreement based on three factors and {PUF for UAVs-assisted} post-disaster
  emergency communication,'' \emph{IEEE Internet Things J.}, vol.~11, no.~11,
  pp. 20\,457--20\,472, Jun. 2024.

\end{thebibliography}

%

\end{document}